\documentclass[twoside]{article}

\usepackage{aistats2020}
\usepackage{natbib}
\usepackage{graphicx}
\usepackage{amsfonts}
\usepackage{amssymb}
\usepackage{bm}
\begin{document}

%

%

\twocolumn[

\aistatstitle{Hierarchical Hidden Markov Jump Processes for Cancer Screening Modeling}

\aistatsauthor{Rui Meng \And Soper Braden \And Jan Nygard, Mari Nygrad \And Herbert Lee}

\aistatsaddress{UCSC \And  LLNL \And Cancer Registry of Norway \And UCSC}

]

\begin{abstract}
Hidden Markov jump processes are an attractive approach for modeling clinical disease progression data because they are explainable and capable of handling both irregularly sampled and noisy data. Most applications in this context consider time-homogeneous models due to their relative computational simplicity. However, the time homogeneous assumption is too strong to accurately model the natural history of many diseases. Moreover, the population at risk is not homogeneous either, since disease exposure and susceptibility can vary considerably. In this paper, we propose a piece-wise stationary transition matrix to explain the heterogeneity in time. We propose a hierarchical structure for the heterogeneity in population, where prior information is considered to deal with unbalanced data. Moreover, an efficient, scalable EM algorithm is proposed for inference. We demonstrate the feasibility and superiority of our model on a cervical cancer screening dataset from the Cancer Registry of Norway. Experiments show that our model outperforms state-of-the-art recurrent neural network models in terms of prediction accuracy and significantly outperforms a standard hidden Markov jump process in generating Kaplan-Meier estimators.
\end{abstract}

\section{Introduction}

Population-based screening programs for identifying undiagnosed individuals have a long history in improving public health. Examples include screening programs for cancer (e.g., cervical, breast, colon), tuberculosis and fetal abnormalities. While the primary objective of such programs is to identify and treat undiagnosed individuals, these cancer screening programs and the population-level, longitudinal datasets associated with them,  present many opportunities for the data-driven, computational sciences. In conjunction with modern analytic and computational techniques, such data have the potential to yield novel insights into the natural history of diseases as well as improving the effectiveness of the screening programs. 

Hidden Markov Models (HMM) are a standard choice for disease progression modeling for at least three reasons. 
First, the underlying disease is represented as an unobserved, latent Markov process. Second, noisy measurements of the disease states are efficiently incorporated as conditional probability distributions in the emission mechanism. Third, any modeling assumptions for a particular application are easily incorporated into the transition probability matrix and emission mechanism.

However, standard HMMs assume that measurements are regularly sampled at discrete intervals which is often not the case in disease screening programs. Measurements are often irregularly sampled because patients come in for screenings at irregular intervals, even if regular screening tests are recommended.
To deal with irregular sampling, Continuous-Time Hidden Markov Models (CTHMM) are often used since they easily handle samples taken at arbitrary time intervals. CTHMMs have been proposed in many applications such as networks \citep{Wei_2002}, medicine \citep{Bureau_2003}, seismology \citep{Lu_2017} and finance \citep{Krishnamurthy_2016}.  Liu summarizes current inference methods for CTHMMs in \cite{Liu_2015} and proposes efficient EM-based learning approaches. 

Because the natural history of many diseases depends heavily on the age of the individual, the time-homogeneous assumption is not valid.  For this reason, time-inhomogeneous HMMs are more appropriate.  Although such models have many appealing theoretical properties according to the Kolmogorov equations \citep{Zeifman_1994}, parameter inference is intractable in most non-trivial cases. For this reason, many inference studies of continuous-time, time-inhomogeneous HMMs (CTIHMMs) in the medical domain depend on inefficient microsimulations \citep{Sonnernberg_1993,Myers_2000,Canfell_2004}

Because of the computational issues, many previous HMM models of disease progression assume that the observations come from a homogeneous population. In large populations, this will typically not be the case. For example, in population-level screening data a large proportion of individuals have benign test results while only a small proportion have abnormal test results. Frailty models are proposed as a common methodology in epidemiological modeling \citep{Amy2010}.

To deal with these difficulties we introduce piece-wise constant transition intensity functions, which allow for tractable parameter inference yet are considerably more flexible in terms of time-inhomogeneity.  We then propose a latent structure (i.e., frailty model) to capture unobserved population heterogeneity in terms of disease exposure and susceptibility.  Specifically, we propose a new hierarchical hidden Markov model for disease progression in which patients are categorized into classes based on risk levels. Due to the expensive cost of the standard EM algorithm inference, we propose an efficient and scalable EM algorithm combining both soft and hard assignment in the E-step and an auto-differentiation based Limited-memory BFGS optimization method in the M-step. 

We apply this model to cervical cancer screening data from the Cancer Registry of Norway. This is a true population-level dataset with over 1.7 million women and more than 10 million screening results. Based on the cervical cancer screening data, our model is demonstrated to have better predictive accuracy compared with state of the art recurrent neural network models, based on AUC (Area Under the Curve) under a binary classification framework. Moreover, our model is significantly better than a simple hidden Markov model by comparing model-generated Kaplan-Meier curves with observed Kaplan-Meier curves.

\section{Related Work}
Longitudinal observation data exist widely, especially in the healthcare area. \cite{Simpson_2013} proposed multiple self-controlled case series to model the multiple drug exposure based on conditional poisson regression. \cite{Bao_2017} extends it from discrete time to continuous time using Hawkes process modeling. Moreover, \cite{Kuang_2017} propose baseline regularization to leverage the diverse health profiles for adverse drug events. It is also a generalized linear model extended from \cite{Kuang_2016}.

For a screening test, the health status is of interest. It is crucial to consider a latent model of health status. The hidden Markov model is the state of the art approach. Most hidden Markov model variants consider only discrete time \citep{Gael_2008,Beal02,Cem_2014}. Continuous time hidden Markov models can handle data at any time stamp \citep{Cox_2017} and therefore are suitable for irregularly-sampled longitudinal data \citep{Bartolonemo_2011,Liu_2013,Wang_2014}. Furthermore, \cite{Liu_2015} summarizes and discusses learning approaches for continuous time hidden Markov models and proposes efficient EM-based learning approaches. Since screening processes significantly depend on a patient's age, our model is based on a CTIHMM, which is discussed in \cite{Sonnernberg_1993}, \cite{Myers_2000} and \cite{Canfell_2004}. 

A modified transition matrix is generally modeled by letting the transition distribution depend on a set of observed covariates or exogenous time-series via a multinomial logistic function \citep{hughes1999non, meligkotsidou2011forecasting, paroli2008bayesian}. Our transition distribution is derived from the jump Markov process using the Kolmogorov equations. 
While the parallelization of EM algorithms for hidden Markov models has been studied \citep{mitchell1995parallel, li2008cut}, to the best of our knowledge, there is no literature on efficient inference for continuous-time, time-inhomogeneous hidden Markov models. In this paper we propose a scalable EM algorithm for the efficient inference of such models, which is much more efficient than a naïve or straight forward implementation of a standard EM algorithm.  


\section{Model}
\begin{figure}
    \centering
    \includegraphics[width=0.8\linewidth]{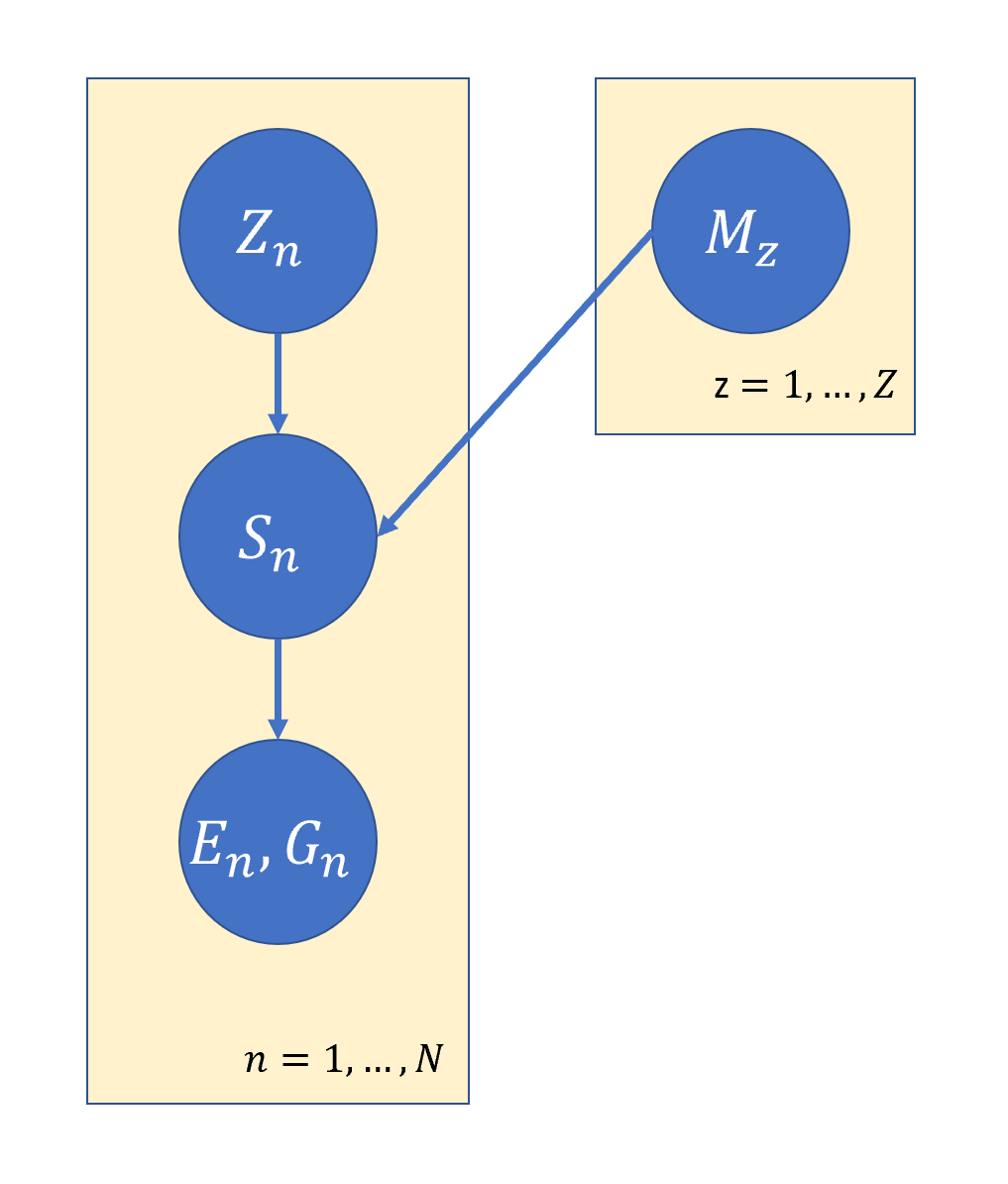}
    \caption{Graphical representation for hierarchical inhomgeneous hidden Markov models.}
    \label{fig:HHMM_gr}
\end{figure}

We propose a hierarchical inhomogeneous HMM (HIHMM) to model disease progression. The hierarchical graphical representation is illustrated in Figure~\ref{fig:HHMM_gr}. In our real world case study, $Z=2$ as we assume all patients come from two risk categories: high disease exposure risk and low disease exposure risk. Each category has its own Markov model shown in Figure~\ref{fig:HHMM}. Details are discussed in Section~\ref{sec:data}. This hierarchical structure of the HIHMM allows for an arbitrary number of latent frailty states, provided relevant Markov models can be ascribed to the disease progression associated with each. 

\subsection{Variables}
Suppose there are $N$ individuals in the screening population.  Let individual $n$ have $T_n$ screening visits at ages $a_1, \ldots, a_{T_n}$. We assume $Z$ categories are considered in the hierarchical frailty structure and introduce the following variables:
\begin{align}
& \textrm{Frailty State (hidden):} z_{n} \in \{1, \ldots, Z\} \nonumber \\
& \textrm{Disease States (hidden):} S_{nt} \in \{1, \ldots, M_{z_n}\} \nonumber \\
& \textrm{Number of screening tests (observable):} E_{ntk} \in \mathbb{N} \nonumber \\
& \textrm{Screening test results (observable):} G_{ntk} \in \mathbb{N}^{L_k} \nonumber 
\end{align}
The underlying disease state of individual $n$ is assumed to evolve according to a continuous-time, time-inhomogeneous Markov process assigned by its latent frailty class indicator $z_n$, where only screening results at specific time stamps with corresponding ages $a_1, \ldots a_{T_n}$ are observable. On the $t$th screening visit of individual $n$, $S_{nt}$ refers to the latent disease state and the visit includes $E_{ntk}$ tests of the $k$th test type and the corresponding results $G_{ntk}$, which is a $L_k$ dimensional vector and the value on $l$th dimension refers to the number of the $l$th grade results. 

\subsection{Model of Disease Progression}
As for the $z$th underlying Markovian disease process, it is parameterized by an $M_z \times M_z$ transition intensity matrix $Q_z$. For the simplicity of notation, we ignore the subscript $z$ in the remainder of this section. The $ij$th element $q_{ij}$ of $Q$ satisfies $q_{ij} \geq 0$ for $i\neq j$ and $q_{ii} = -\sum_{i\neq j}q_{ij}$.    
The time spent in state $i$ is exponentially distributed with rate $-q_{ii}$.  Given that a transition occurs from state $i$,  the probability of transitioning to state $j$ is $\frac{q_{ij}}{q_i}$ where $q_i = \sum_{i\neq j}q_{ij}$.
When $Q$ is invariant for time $t$ the model is homogeneous, otherwise the model is inhomegeneous. 

\subsubsection{Homogeneous Markov Model} \label{sec:homogeneous}
For a homogeneous Markov process, we assume the initial state at $t_0$ is known, $p(S(t_1)) = 1$. We let $\bm t' = (t'_1, \ldots, t'_{T'})$ refer to the underlying transition timestamps and let $\bm O = (O_1, \ldots, O_T)$ denote observations at time $\bm t = (t_1, \ldots, t_T)$. Then the complete likelihood (CL) is given by 
\begin{align}
& \mathrm{CL} = \prod_{i = 1}^{T'}(q_{S(t'_i),S(t'_{i+1})}/q_{S(t'_i)})q_{S(t'_i)}e^{-q_{S(t'_i)}\vartriangle_i} \prod_{j = 1}^Tp(O_j|S(t_j)) \nonumber\\
& \qquad = \prod_{i = 1}^{M}\left(e^{-q_i\tau_i} \prod_{j\neq i} q_{ij}^{n_{ij}}\right)\prod_{j=1}^Tp(O_j|S(t_j))\,,
\label{CL}
\end{align}
where $\tilde{\vartriangle}_i = \tilde{t}_{i+1} - \tilde{t}_i$ and $n_{ij}$ denotes the number of times the state changes from state $i$ to state $j$ during the whole process and $\tau_i$ denotes the duration that the process stays in state $i$.
Since the underlying transition timestamps $\bm t'$ are not observable, the marginalized complete likelihood (MCL) is derived by marginalizing all $\bm t'$ as
\begin{align*}
\mathrm{MCL} = \prod_{i = 1}^{T-1}P(\vartriangle_i)_{S(t_i), S(t_{i+1})}\prod_{j = 1}^T p(O_j|S(t_j)) \,,
\end{align*}
where $\vartriangle_i = t_{i+1} - t_i$ and $P(\vartriangle_i) = e^{Q \vartriangle_i }$ is the transition probability matrix from time $t_i$ to time $t_{i+1}$.  

\subsubsection{Inhomogeneous Markov Model} \label{sec:inhomogeneous}

An inhomogeneous Markov process drops the time invariance assumption of $Q$ by allowing $Q$ to be a function of $t$. CL then becomes intractable, because the time spent in state $i$ no longer follows an exponential distribution. An alternative approach is to consider the MCL. The only difference in the expression of MCLs between homogeneity and inhomogeneity is the computation of the transition matrix $P([t_i, t_{i+1}])$ from time $t_i$ to time $t_{i+1}$ for $i = 1, \ldots, T - 1$. For the inhomogeneous model, $P([t_i, t_{i+1}]) = \exp\{{\int _{t_i}^{t_{i+1}}Q(t)dt}\}$.

The transition intensity function $Q(t)$ can be modeled by any parametric model, but the computation of the matrix exponential $\exp\{{\int _{t_i}^{t_{i+1}}Q(t)dt}\}$ may be prohibitively expensive, even taking  advantage of numerical computational methods. To ease this computational burden, we propose a piecewise constant transition intensity matrix $Q$, where each element $q_{ij}$ is a piecewise constant function of time. Specifically, we partition time into $I$ disjoint intervals covering the range of observable time.  We then have a set of disjoint partitions $\mathcal{A} = \{ A_i\}_{i=1}^{I}$. Each transition intensity function $q_{ij}$ is a piecewise constant function via the defined partition $\mathcal{A}$, denoted by $q_{ij}(t) = \sum_{k = 1}^{I}q_{ijk} \bm 1_{A_k}(t)$, where $\bm 1(\cdot)$ is an indicator function and $q_{ij\ell} \geq 0$. In this case, the inhomogeneous Markov process can be treated as a combination of several continuous-time homogeneous Markov processes, and the transition probability matrix $Q$ can be computed as a product of transition probability matrices with respect to their corresponding partitions.

\subsection{Hierarchical Model}

Due to the significant population heterogeneity related to disease exposure risk, we propose a hierarchical model as follows. Let $\bm \psi = (\bm \psi_1, \ldots, \bm \psi_Z)$ denote all model parameters and $\bm \psi_z$ be parameters for model $z$. Then the hierarchical model is given by
\begin{eqnarray*}
	\bm O_n & \sim & \mathcal{M}_{z_n}(\bm \psi_{z_n}, \bm \theta_n)\,, \\
	z_n & \sim & \mathrm{Cat}(\bm p)\,, 
\end{eqnarray*}
where $\bm \theta_n$ denotes all covariates for individual $n$. An informative prior of the model indicator $z_n$ is proposed as a categorical distribution with hyper-parameters $\bm p$, which is used to provide expert knowledge of the model assignment. This prior contributes to reasonable model inference, especially when screening data are highly unbalanced in terms of latent class membership.    Figure~\ref{fig:HHMM} shows the case where $Z=2$ and index $z_n$ has a Bernoulli prior with a parameter $p$, i.e.,\ $z_n \sim \mathrm{Ber}(p)$.

\section{Inference}
Due to the latent characteristics of both model indices and patient states, the expectation maximization (EM) approach is considered. Our EM alogrithm employs the true conditional posterior for model index and the pseudo-conditional posterior for states in the E-step. Both hard-assignment and soft assignment approaches are studied in literature but they are not combined. We balance the advantages of both methods for inference. Specifically, considering the heterogeneity of our model, we marginalize the latent transition timestamps in our inference. We decompose the joint posterior distribution as $p(z_n, \bm S_n| -) = p(z_n| - )p(\bm S_n| z_n, -)$, where $-$ denotes all other parameters and use soft assignment for $p(z_n| -)$ and hard assignment for $p(\bm S_n| z_n, -)$, because the computation of $p(\bm S_n|z_n, -)$ is prohibitively expensive. For simplicity, we ignore covariates $\bm{\theta}_n$ in the remainder of this section. 

The recursive procedures are given as follows:
\begin{itemize}
	\item Given previous estimates $\bm \psi^{(t-1)}$, compute the conditional posterior distribution of $z_n$:
	\begin{align}
	p(z_n|\bm O_n, \bm \psi^{(t-1)}) & \propto \mathrm{Cat}(z_n|\bm p)p(\bm O_n| z_n, \bm\psi^{(t-1)}_{z_n}) \nonumber \\
	& \sim \mathrm{Cat}(\tilde{\bm p}_n)\,,
	\label{Pos_I}
	\end{align}
	where $\tilde{p}_{nk} = \frac{p_k p(\bm O_n| z_n = k, \bm{\psi}^{(t-1)}_k)}{\sum_{z = 1}^{Z}p_z p(\bm O_n | z_n = z, \bm\psi^{(t-1)}_z)}$ for $k = 1, \ldots, Z$ and $p(\bm O_n | z, \bm \psi_{z})$ is accessible through the forward-filter backward-sample algorithm (FFBS), which is a sequential Monte Carlo approach first proposed in \cite{Kitagawa_1987}.
	\item Update the optimal state sequence $\bm S_n$ given corresponding observations $\bm O_n$ and model indicator $z$ using the Viterbi algorithm \citep{Forney_1973}: 
	\begin{align}
	\bm S^{(t)}_{nz} = \mathrm{Viterbi}(\bm O_n, \bm \psi_z)\,.
	\label{Pos_S}
	\end{align}
	\item Maximize the expected marginal complete log-likelihood (EMCLL) with respect to $\bm \psi$ by
	\begin{align}
	\bm \psi^{(t)} & = \arg\max\limits_{\bm{\psi}}\sum_{n=1}^{N}E_{z_n, \bm S_n}(\ell(\bm \psi| \bm O_n, z_n, \bm S_n)| \bm O_n, \bm \psi^{(t-1)}) \nonumber \\ 
	& =   \arg\max\limits_{\bm{\psi}} \sum_{n = 1}^{N}\sum_{z = 0}^{Z}p(z_n = z| \bm O_n, \bm \psi^{(t-1)}) \nonumber \\
	& \qquad \left(\log p_z + \log p(\bm S_n | z, \bm \psi) + \log p(\bm O_n| \bm S_n, \bm \psi) \right)\,.
	\label{M_step}
	\end{align}
\end{itemize}

Since population screening datasets can contain millions of records, direct inference may be prohibitively expensive and more scalable approaches are necessary. We scale our EM algorithm by parallelizing the inference across observations using $\tilde{N}$ clusters, $\{C_n\}_{n=1}^{\tilde{N}}$, in three parts. We first compute the conditional posterior distribution of $z_n$ in each cluster using (\ref{Pos_I}). The time complexity for each cluster is $O(|C_n|ZM^2T)$ \citep{KHREICH_2010}. We then compute the optimal state sequences in each cluster $C_n$ using (\ref{Pos_S}) with the same time complexity $O(|C_n|ZM^2T)$ \citep{Arturs_2016}. Finally, we compute the gradients in each cluster then reduce all local gradients to global gradients for the optimization in the M-step. In detail the EMCLL is rewritten as
\begin{align*}
	& \sum_{n=1}^{N}E_{z_n,\bm S_n}(\ell(\bm \psi| \bm O_n, z_n, \bm S_n)| \bm O_n, \bm \psi^{(t-1)}) \\ 
	= & \sum_{\tilde{n}=1}^{\tilde{N}}\sum_{n\in C_{\tilde{n}}}E_{z_n,\bm S_n}(\ell(\bm \psi| \bm O_n, z_n, \bm S_n)| \bm O_n, \bm \psi^{(t-1)}).
\end{align*} 
Taking gradient on both sizes, we have
\begin{align*}
	& \frac{\partial}{\partial \bm{\psi}}\sum_{n=1}^{N}E_{z_n,\bm s_n}(\ell(\bm \psi| \bm O_n, z_n, \bm S_n)| \bm O_n, \bm \psi^{(t-1)}) \\
	= & \sum_{\tilde{n}=1}^{\tilde{N}} \frac{\partial}{\partial \bm{\psi}}\sum_{n\in C_{\tilde{n}}}E_{z_n,\bm S_n}(\ell(\bm \psi| \bm O_n, z_n, \bm S_n)| \bm O_n, \bm \psi^{(t-1)}).
\end{align*}
Automatic differentiation (AD) \citep{Baydin_2018} is utilized to compute the gradients in each cluster. Summing over all clusters, we get the gradient of the EMCLL. Using this gradient we adapt the Limited-Memory BFGS \citep{Liu_1989} algorithm to estimate $\bm \psi$. Analytically computing the complexity of the L-BFGS algorithm for each cluster is intractable, but the paralleled algorithm increases inference speed around $\tilde{N}$ times.

\section{Experimental Results}\label{sec:data}
The HIHMM is demonstrated on a true population-level cervical cancer screening test dataset from the Cancer Registry of Norway. Data used in the analyses will be available on request from the Cancer Registry of Norway, given legal basis according to the GDPR. 
This dataset contains 1.7 million patients' screening testing records. Each patient has a censored observation at the last time stamp $t_{c}$, denoted by $O_{c}$, which indicates whether the woman is dead or alive at time $t_c$. Each patient has treatment indices to show when and how many treatments occurred, and results of screening tests for each of cytology, histology and hpv. Cytology and histology have four levels of outcomes while hpv has two levels.

\begin{figure}[ht!]
	\centering
	\includegraphics[width = 0.3\textwidth]{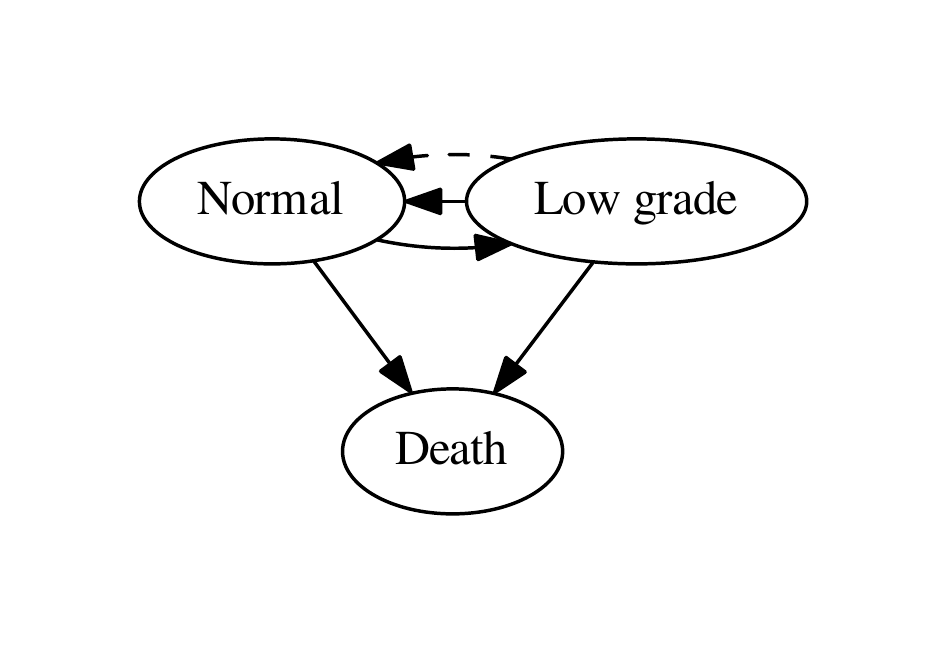}
	\includegraphics[width = 0.4\textwidth]{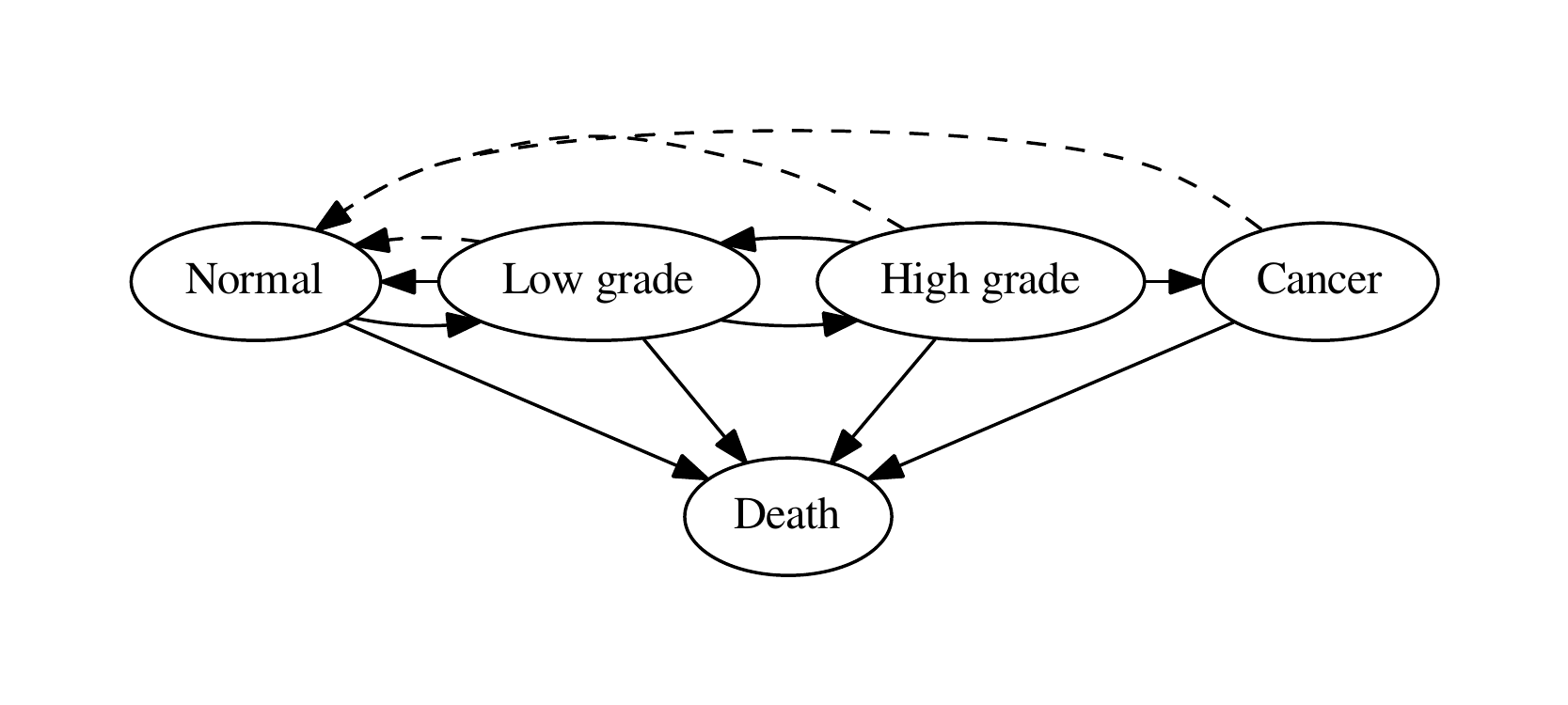}
	\caption{Transition structure of model $\mathcal{M}_0$ and $\mathcal{M}_1$. Solid lines denote the intensity transition while dashed lines denote that any state comes back to the normal state once treatment is completed.}
	\label{fig:HHMM}
\end{figure}

We set $Z = 2$ and the model processes are displayed in Figure~\ref{fig:HHMM}. For model $z$, the initial state is modeled as $S_{z1}|a_{1} \sim \mathrm{Cat}(\bm \pi_z(\mathcal{A}, a_0))$ and $\bm \pi_{zi} \sim \mathrm{Dir}(\bm \alpha_{zi})$, where $a_1$ denotes the age at the first screening test and $\mathcal{A}$ is a disjoint partition of observable ages, $\bm \pi_z(\mathcal{A}, a) = \bm \pi_{zi}$ if and only if $a \in \mathcal{A}_{i}$, and $\bm{\alpha}_{z\ell} \in \mathbb{R}^{+M_z}$.

The observations $\bm O$ have two levels: the number of screening tests $\bm E$ and the results of screening tests $\bm G$. Omitting the subscripts $n$ and $t$, given state $s$, observations are modeled as  
\begin{eqnarray*}
	E_k & \sim & \mathrm{Poisson}(\eta_{sk})\,, \nonumber \\
	{\bm G}_k|E_k & \sim & \mathrm{Multinomial}(E_k, \tilde{\bm\pi}_{sk})\,, \nonumber \\
	\tilde{\bm\pi}_{sk} & \sim & \mathrm{Dir}(\tilde{\bm\alpha}_{sk}) \,,
\end{eqnarray*}
where 
$\tilde{\bm \alpha}_{sk} \in \mathbb{R}^{+L_k}$ are hyper-parameters for observation model. The censored observations (dead or alive) are modeled by
\begin{eqnarray*}
	p(O_{c}|S_T) = \begin{cases}
		P(t_T, t_{c})_{S_T, \text{death}} & \text{ if } O_{c} = \mathrm{death}\,,\\
		1-P(t_T, t_{c})_{S_T, \text{death}} &   \text{ if }  O_{c} \neq \mathrm{death}\,.
	\end{cases}
\end{eqnarray*}

We choose the age partition as $\mathcal{A}$ as $[0, 23)$, $[23,30)$, $[30,60)$ and $[60, \infty)$. More details are available in the supplementary material, including treatment modeling and inference
, age partition selection, model training and optimization, which are all available in the Appendix.

In the proposed learning approach, we set the number of EM iterations at $N_{\mathrm{EM}} = 100$, and in the Limited-memory BFGS (L-BFGS) approach we set the number of optimization iterations as $N_{\mathrm{L-BFGS}} = 8$.  Automatic differentiation is implemented using the autograd package \citep{Maclaurin_2016} in Python.

\subsection{Model Comparison}
We randomly select 80000 patients' records for training and select another 20000 records for testing. The goal is to predict the status at the last visit. Specifically, if a patient has at least one result whose level is greater than 1, then the status is defined as high risk denoted as $1$. Otherwise, the status is defined as low risk denoted as $0$. Thus, the problem is defined as a binary classification problem. 

The prediction procedure is defined as follows. After model training, let model parameter estimates be $\hat{\bm \psi}$. Given new patient historical records $\bm O^*$, compute the predictive distribution of model index $p(z^*|\bm O^*, \hat{\bm \psi})$. Next, given model index $z$, compute the predictive distribution of the state at the second to last visit
$p(S^*_{T-1}|z, \bm O^*, \hat{\bm \psi})$ derived from FFBS. Then the predictive distribution of the state of the last visit is 
$p(S^*_{T}|\bm O^*, \hat{\bm \psi}) \sum_{z} p(z^* = z|\bm O^*, \hat{\bm \psi}) p(S^*_{T}|z, \bm O^*, \hat{\bm \psi})$ and the predictive distribution of screening test results is $p(\bm G^*_{T}| \bm O^*, \bm E^*_T, \hat{\bm \psi}) = \sum_s p(S^*_{T} = s|\bm O^*, \hat{\bm \psi}) p(\bm G^*_{T}| S_{T}^* = s, \bm E_{T}^*, \hat{\bm \psi})$. Finally, the predictive distribution of the last status is $G^* \sim \mathrm{Ber}\left( p^* \right)$, where $p^* = p\left(\sum_{i = 0}^{1}\sum_{j = 2}^3 \bm G_{T}^*[i,j] \geq = 1 |  \bm O^*, \bm E^*_T, \hat{\bm \psi}\right)$. and it is estimated by $\hat{G}^* = \begin{cases}
1 & p^* \geq 0.5 \\
0 & \mathrm{otherwise} \\
\end{cases}$. And the similar procedures are available for the CTIHMM.

Recurrent neural network models (RNN) have been found to perform well with variable-length time series and capture the temporal correlation well, because of their flexibility and lack of the Markov property. Variants of RNNs have been proposed to better balance memory needs and new features. \cite{Hochreiter_1997, Cho_2014} propose the concept of a Long Short-Term Memory (LSTM) architecture, with variants utilized in handwriting recognition \citep{Doestsch_2014}, language modeling \citep{Stephen_2017}, and video data \citep{Zhang_2016}. \cite{Chung_2014} proposes the gated recurrent neural network (GRU) as another state of the art RNN model. We compare RNN models with our proposed model on the same prediction task to illustrate that our model with interpretability outperforms RNNs on those irregular samples. In details, each patient's record is modeled as one time-series and the features at each visit includes patient's age, patient's screening result and patient's treatment indicator. The screening result of patient $n$ at the $t$th visit is $\vec{\bm G}_{n,t}$. The treatment indicator is equal to 1 if and only if the patient has accepted treatment. LSTM \citep{Cho_2014}, stacked LSTM \citep{Dyer_2015} and GRU \citep{Chung_2014} with different sizes are implemented for model comparison. We propose small and large size of models where small model refers to layer size $16$ while large model refers to layer size $64$.  For stacked LSTM, two LSTMs are stacked. In the HIHMM, through cross validation, the model prior is set as $p = 0.001$ and we set 100 iterations in EM algorithm. In addition, we set 10 iterations in EM algorithm as a fast inference. In the CTIHMM, we also set 100 iterations in EM algorithm. We summarize both prediction results and training time in Table~\ref{Model_Comparison}. It shows our model outperforms state of the art methods overall on the set of criteria: Area Under the Curve (AUC), F1 value (F1), Average Precision (AP) and Recall (R). Those metrics related to recall score are important in clinic diagnosis, it is better to mis-classify low-risk patient rather than high-risk patient. Our model also have competitive running time compared with neural network models.

\begin{table*}[ht!]
	\centering
	\caption{Model prediction for the status of the last visit in terms of Accuracy (ACC), Area Under The Curve(AUC), F1, Average Precision (AP), Precision (P), Recall (R).}
	\begin{tabular}{|c|c|c|c|c|c|c|c|}
		\hline
		Method & ACC & AUC & F1 & AP & P & R & training time (h) \\
		\hline
		LSTM (small) & 0.9905 & 0.4939 & 0.0000 & 0.0095 & 0.0000 & 0.0000 & 1.11 \\
		\hline
		LSTM (large) & 0.9925 & 0.8563 & 0.4275 & 0.2359 & 0.7778 & 0.2947 & 3.38 \\
		\hline
		stacked LSTM (small) & 0.9914 & 0.8561 & 0.2773 & 0.1273 & 0.6875 & 0.1737 & 2.26 \\
		\hline
		stacked LSTM (large) & \textbf{0.9926} & 0.8573 & 0.4335 & 0.2409 & \textbf{0.7808} & 0.3000 & 7.5 \\
		\hline
		GRU (small) & 0.9920 & 0.8379 & 0.4089 & 0.2083 & 0.6962 & 0.2895 & 0.79 \\
		\hline
		GRU (large) & 0.9921 & 0.8678 & 0.4207 & 0.2178 & 0.7037 & 0.3000 & 2.30 \\
		\hline
		CTIHMM & 0.9910 & 0.9128 & 0.3466 & 0.1465 & 0.5517 & 0.2526 & 3.61 \\
		\hline
		HIHMM & 0.9914 & 0.9190 & \textbf{0.5210} & \textbf{0.2774} & 0.5589 & \textbf{0.4895} & 6.97 \\
		\hline
		HIHMM (fast) & 0.9912 & \textbf{0.9268} & 0.5014 & 0.2583 & 0.5466 & 0.4632 & 0.90 \\
		\hline
	\end{tabular}
	\label{Model_Comparison}
\end{table*}

\subsection{Model Validation}
Based on epidemiological studies, the incidence rate of cervical cancer can provide us with guidance on how to choose priors for the frailty rate of a given population \citep{Bray2018}.
An informative prior is indeed preferable here, so we set a conservative model index prior $p = 0.2$. We note that there is always a trade off between precision and recall, and $p = 0.2$ provides a reasonable balance based on model comparison results. We present two types of results on population-level data. First we present the MLEs for all model parameters along with bootstrapped standard deviations. Second we perform model validation using Kaplan-Meier estimators as suggested in \cite{titman_general_2008} and predictive accuracy via proposed average posterior predictive probability.

We randomly divide all data into clusters such that each cluster has 100 individual observation sequences. Using a bootstrap technique, we randomly select $2400$ clusters with replacement for model inference. We independently repeat the same inference on different selections 5 times. The mean and standard deviation of all parameter estimates are discussed in Appendix.

For model validation we randomly select $2400$ clusters of data in which each cluster has 100 individual sequences of observations. We implement both the HIHMM and the CTIHMM for the same dataset. We follow the method proposed in \cite{titman_general_2008} that utilizes Kaplan-Meier estimators to validate continuous-time HMMs. Kaplan-Meier estimators are defined according to the definition of a failure, or time-to-event. In multi-state models different failures can be defined depending on which features of the model and data are of interest. Here we define failure as the first observation of a high-risk or cancer test result directly following an initial normal or low-grade test result.  Accurately predicting this time-to-event is of practical importance because clinical intervention is only possible in the high-grade state.  Treating patients at this stage is critical to preventing precancerous lesions from progressing to cervical cancer. 

\begin{figure}
	\centering
    \includegraphics[width = 0.45\textwidth]{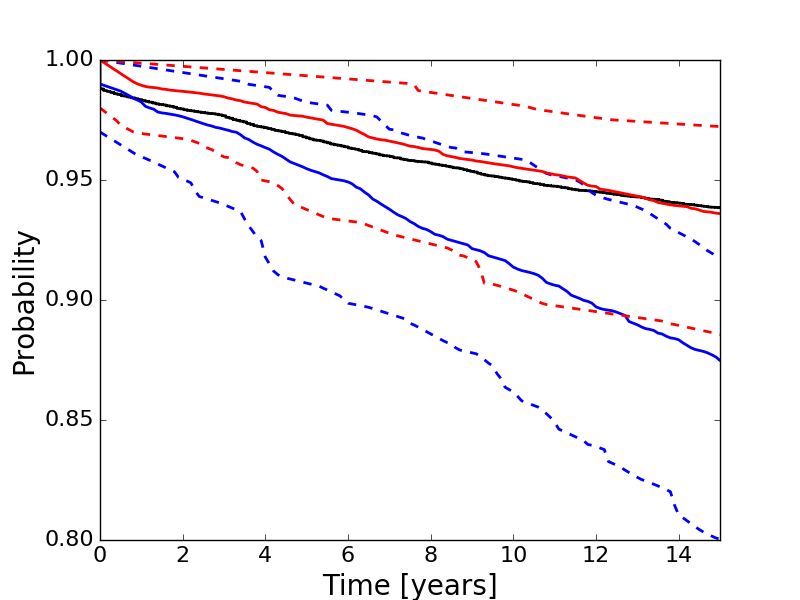}
    \includegraphics[width = 0.45\textwidth]{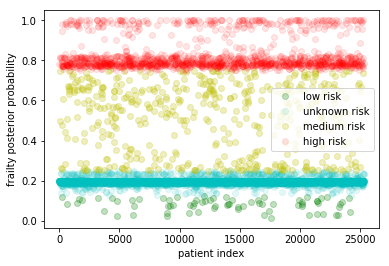}
	\caption{Top panel: Empirical Kaplan-Meier curve (black) and simulated Kaplan-Meier curves, which are summarized using the 95\% credible interval (dashed lines) and the median (solid lines), from the CTIHMM (blue) and HIHMM (red). Bottom panel: Posterior probabilities of belonging to the frailty class for each individual from a test set. Risk stratification is possible by thresholding the probabilities. Threshold probabilities in this example are $(0, 0.125,0.25, 0.75,1)$. Color indicates falling between two probability thresholds.}
	\label{KM_PF}
\end{figure}

The empirical Kaplan-Meier estimator is an important criterion because it measures prediction on the whole process rather than only the last visit, and it is defined as 
$\hat{S}(t) = \prod_{i: t_i\leq t}\left(1 - d_i/n_i\right)$,
where $t_i$ is a time when at least one failure is observed, $d_i$ is the number of failures that occurred at time $t_i$, and $n_i$ is the number of individuals known to have survived up to time $t_i$. We randomly choose $24000$ records to generate an empirical Kaplan-Meier estimator according to our definition of a failure. We generate Kaplan-Meier estimates by simulating $100$ sequences from both the CTIHMM and HIHMM and repeat this $100$ times. Figure~\ref{KM_PF} shows the empirical Kaplan-Meier curve in black, simulated Kaplan-Meier curves from the CTIHMM in blue, and simulated Kaplan-Meier curves from the HIHMM. As for the simulated Kaplan-Meier curves, solid lines denote the median curve and dashed lines denote the 95\% credible intervals based on the 100 replications. The results show that the empirical Kaplan-Meier curve is always near the median and within the 95\% credible intervals generated by the HIHMM. This is not the case with the CTIHMM. In this sense the HIHMM outperforms the CTIHMM in an important clinical metric.

On the other hand, the HIHMM has a relatively high Kaplan-Meier estimate at time $0$ because the informative prior $p = 0.2$ is relatively small. This has the effect of driving simulated patients to more likely be in the low-risk model $\mathcal{M_0}$ at the initial time. Moreover, these patients are more likely to stay at the normal state for longer. However, the trend of the median curve from the HIHMM more closely tracks that of the empirical Kaplan-Meier curve, compared with the trend of the median curve from the CTIHMM. This suggests that the HIHMM models disease progression better than the CTIHMM. The Kaplan-Meier curves simulated from the CTIHMM are always underestimated.

We also propose average posterior predictive probability for Cytology, Histology and HPV at the last visit given the screening tests as a quantitative measurement for model validation. Specifically, for each testing patient, its posterior predictive probabilities are $p(O_{T, \mathrm{test}}^*| O_{-T}^*, \hat{\bm \psi}, E_{T}^*)$ where tests include Cytology, Histology and HPV and model parameters $\hat{\bm \psi}$ are estimated via model training. We take 240000 patients for training and 20000 patients for testing and model validation results for CTIHMM and HIHMM are summarized in Table~\ref{tab:model validationa}.

\begin{table}[ht!]
    \centering
    \begin{tabular}{|c|c|c|c|}
         \hline
         Model & Cytology & Histology & HPV \\
         \hline
         CTIHMM & 0.9563 & 0.7597 & 0.6550 \\
         \hline
         HIHMM & \textbf{0.9571} & \textbf{0.7613} & \textbf{0.7010} \\
         \hline
    \end{tabular}
    \caption{Average posterior predictive probabilities of Cytology, Histology and HPV for CTIHMM and HIHMM models}
    \label{tab:model validationa}
\end{table}

\section{Conclusion and Discussion}
One of the possible applications of the HIHMM in the context of population-based screening programs is risk stratification of the population.  The latent random variable $z_n$ is an indicator of belonging to a frail class in the population.  Given the learned model parameters $\bm \psi$  it is possible to compute the posterior probability of belonging to the frailty class for individual women.  In other words, given an observed sequence of test results ${\bm O}_n$ and model parameters  $\bm \psi$,  the posterior predictive distribution $p(z_n | \bm \psi, {\bm O}_n)$ is of interest.  This parameter gives a measure of the likelihood of an individual to be at risk of developing cervical cancer conditioned on their observed test results. Such information could be used to more efficiently screen a population by avoiding the over screening of women at low-risk and the under screening of women at high-risk.

Examples of these posterior probabilities are shown in Fig \ref{KM_PF}. For illustration purposes, we have chosen risk thresholds of $\{0.125,0.25, 0.75\}$ with the following interpretation.
\begin{align*}
0 \leq p(z_n | \bm \psi, {\bm O}_n) < 0.125 &\implies \text{ low-risk  } \\
0.125 \leq p(z_n | \bm \psi, {\bm O}_n)  < 0.25 &\implies \text{ unknown risk  }\\
0.25 \leq p(z_n | \bm \psi, {\bm O}_n)  < 0.75      &\implies \text{ medium-risk }\\
0.75 <  p(z_n | \bm \psi, {\bm O}_n) \leq 1 &\implies \text{ high-risk } 
\end{align*}

Two main clusters are apparent in the data corresponding to unknown risk and high risk. The unknown risk cluster is those patients close to the prior probability of $20\%$.  These patients lack sufficient observations to make an informed decision about their risk profile.  This suggests these patients should be followed up with the standard screening protocol.  The high risk cluster is those patients who are more likely to be in a high-grade state.  This suggests these patients may require immediate follow up.  The two smaller clusters of low risk and medium risk are comprised of patients that may require decreased or increased screening frequencies, respectively, relative to the standard screening protocol.

In summary, this paper has made the following contributions:
\begin{itemize}
	\item We model treatment effects in CTIHMM and make CTIHMM inference possible for population-level datasets by using piece-wise constant intensity functions and deriving a scalable EM-based inference algorithm.
	\item We put a hierarchical structure over the CTIHMM to explain population heterogeneity in terms of frailty but share the same states and their emission probability, which makes the model more practical, resulting in our HIHMM.
	\item We utilize prior distributions in the model to achieve more accurate estimates when dealing with imbalanced data.    
	\item We perform full model inference and prediction on subset of a cancer screening dataset and show that our model outperforms comparators on the prediction task.
	\item We perform full model inference on a cancer screening dataset and show that modeling population heterogeneity improves performance in terms of Kaplan-Meier estimators and proposed average posterior predictive probability. 
	\item We illustrate how the model may be used to better inform public health professionals by providing a risk stratification mechanism. 
\end{itemize}

\bibliographystyle{apalike}
\bibliography{ref}

\end{document}